\documentclass[aps,prd,superscriptaddress,twocolumn,nofootinbib]{revtex4-2}
\usepackage[utf8]{inputenc}
\usepackage{dsfont}
\usepackage{amsfonts}
\usepackage{amsmath}
\allowdisplaybreaks[4]        
\usepackage{amssymb}
\usepackage{euscript}     
\usepackage{braket}
\usepackage{starfont}
\usepackage{color,soul}         
\usepackage{tensor}        
\usepackage{amsthm}
\usepackage{graphicx}
\usepackage{slashed}
\usepackage{leftidx}
\usepackage{subfigure}
\usepackage{bbm}
\definecolor{outerspace}{rgb}{0.25, 0.29, 0.3}
\definecolor{scarlet}{rgb}{1.0, 0.13, 0.0}
\usepackage[header,title,page,titletoc]{appendix}  
\definecolor{princetonorange}{rgb}{1.0, 0.56, 0.0}
\definecolor{WildStrawberry}{rgb}{1.0, 0.26, 0.64}
\definecolor{rossocorsa}{rgb}{0.83, 0.0, 0.0}
\definecolor{navyblue}{rgb}{0.0, 0.0, 0.5}
\usepackage{float}
\usepackage[paper=letterpaper,margin=1in]{geometry}
\usepackage{mathtools}

\DeclareMathAlphabet{\pazocal}{OMS}{zplm}{m}{n}
\newcommand{\req}[1]{(\ref{#1})} 

\newcommand{\bea}{\begin{eqnarray}}

\newcommand{\eea}{\end{eqnarray}}
\newcommand{\ba}{\begin{eqnarray}}
\newcommand{\ea}{\end{eqnarray}}

\newcommand{\be}{\begin{equation}}
\newcommand{\ee}{\end{equation} }
\newcommand{\beqa}{\begin{eqnarray}}
\newcommand{\eeqa}{\end{eqnarray}}
\newcommand{\beqar}{\begin{eqnarray*}}
\newcommand{\eeqar}{\end{eqnarray*}}

\renewcommand{\req}[1]{(\ref{#1})}

\makeatletter
\newcommand{\dal}{\mathop{\mathpalette\dal@\relax}}
\newcommand{\dal@}[2]{%
  \begingroup
  \sbox\z@{$\m@th#1\square$}%
  \dimen0=\fontdimen8
    \ifx#1\displaystyle\textfont\else
    \ifx#1\textstyle\textfont\else
    \ifx#1\scriptstyle\scriptfont\else
    \scriptscriptfont\fi\fi\fi3
  \makebox[\wd\z@]{%
    \hbox to \ht\z@{%
      \vrule width \dimen0
      \kern-\dimen0
      \vbox to \ht\z@{
        \hrule height \dimen0 width \ht\z@
        \vss
        \hrule height 2\dimen0
      }%
      \kern-2.5\dimen0
      \vrule width 2.5\dimen0
    }%
  }%
  \endgroup
}
\makeatother

\usepackage{hyperref}
\hypersetup{
    colorlinks,
    citecolor=navyblue,
    filecolor=navyblue,
    linkcolor=navyblue,
    urlcolor=navyblue
}

\begin{document}

\title{Inconsistency of modified gravity in cosmology}

\author{Pablo A. Cano}
\email{pablo.cano@icc.ub.edu}
\affiliation{Departament de F\'isica Qu\`antica i Astrof\'isica, Institut de Ci\`encies del Cosmos\\
 Universitat de Barcelona, Mart\'i i Franqu\`es 1, E-08028 Barcelona, Spain }


\begin{abstract}
We show that there is a fundamental flaw in the application of modified gravity theories in cosmology, taking $f(R)$ gravity as a paradigmatic example. This theory contains a scalar degree of freedom that couples to the matter stress-energy tensor but not to gravitational energy. However, when applied to cosmology this theory is unable to distinguish between gravitational and non-gravitational energy.  Hence the cosmological version of the theory does not coincide with its own macroscopic average, and we show that this leads to order-one discrepancies. We argue that the same inconsistency is common to many other modified gravity theories with extra degrees of freedom. 
Our results put into question whether these theories can make sense as the cosmological average of a fundamental theory, hence challenging their physical significance. 

\end{abstract}
\maketitle

Our understanding of the large scale structure of the universe is based on the cosmological principle: the idea that, on very large scales, the universe is homogeneous and isotropic. This allows one to describe the geometry of the universe in terms of Friedmann-Lema\^itre-Robertson-Walker (FLRW) metrics, 
\begin{equation}\label{FLRW}
d\hat{s}^2=-dt^2+a(t)^2d\Sigma^2\, ,
\end{equation}
where the spatial metric $d\Sigma^2$ is maximally symmetric and has constant curvature.  The application of Einstein's equations to these cosmological metrics then yields the Friedmann equations for the scale factor $a(t)$, which determine the evolution of the universe. Furthermore, Einstein's equations also rule the behavior of perturbations on top of this background and are able to explain the growth of structure and the inhomogeneities in the CMB power spectrum among other observed phenomena.

But a fundamental question is whether one can actually apply Einstein's equations to the cosmological metrics in the first place. In fact, the cosmological metric $\hat{g}_{\mu\nu}$ (denoted with a hat) is not the same as the fundamental metric $g_{\mu\nu}$
\begin{equation}
\hat{g}_{\mu\nu}\neq g_{\mu\nu}\, ,
\end{equation}
because the spacetime is actually not homogeneous at short scales. Rather, $\hat{g}_{\mu\nu}$ must be regarded as an average metric over very long distances. Thus, even if $g_{\mu\nu}$ satisfies Einstein's equations, one cannot immediately conclude that $\hat{g}_{\mu\nu}$ also satisfies them, 
\begin{equation}
G_{\mu\nu}=8\pi G\, T_{\mu\nu} \quad \Rightarrow \quad \hat{G}_{\mu\nu}=\,\, ?
\end{equation}
where $G_{\mu\nu}$ and $\hat{G}_{\mu\nu}$ are the Einstein tensors of each of the metrics. 
 Indeed, this is a challenging question known as the averaging problem in cosmology \cite{Zalaletdinov:1996aj,Boersma:1997yt,Coley:2005ei,Buchert:2007ik,Green:2010qy,Green:2014aga,Bolejko:2016qku}. 
As it turns out, the analysis of the ``averaged'' or ``macroscopic'' Einstein field equations (\textit{i.e.}, the equations satisfied by the cosmological metric) reveals that these are different from the original Einstein field equations one starts with due to the nonlinearity of the Einstein tensor. 
Although this difference --- due to the backreaction of inhomogeneities --- has been argued to be small in our universe \cite{Green:2014aga}, one should always bear in mind that the equations of motion satisfied by the cosmological metric do not take the same form as the fundamental equations of motion of our theory.  

Now, these considerations have so far been largely unexplored in the case of modified gravity theories --- see \cite{Vitagliano:2009zy, BeltranJimenez:2013zfo, Preston:2014tua} for a few exceptions though. 
Theories such as $f(R)$ gravity \cite{Sotiriou:2008rp}, Horndeski \cite{Horndeski:1974wa} and other extensions of GR --- see \cite{Deffayet:2009wt,Zumalacarregui:2013pma,Gleyzes:2014dya,Langlois:2015cwa,BenAchour:2016fzp} for a few examples --- have been extensively used in the cosmological context in order to explain the accelerated expansion of the universe, dark matter or inflation --- see the reviews \cite{Clifton:2011jh,Capozziello:2011et}. Thousands of scientific papers routinely utilize these theories, but however,  basic questions about the meaning and validity of these theories are often forgotten.

In the light of our previous discussion, the following question is pertinent: if one of these theories is assumed to correspond to the fundamental description of gravity, can one still apply the equations of motion of that theory to cosmological metrics? Is this a good approximation to the result of averaging the theory?
In this note, we argue that the answer is no, as the direct application of these theories to cosmological metrics leads to a different prediction with respect to the one obtained from averaging the theory over large distances. 

Importantly, this difference can be an order-one effect, meaning that the effect of averaging is as important as the effect of the deviations to GR that we are trying to implement with these modified gravity theories. Hence, the naive application of these theories in cosmology would be inconsistent.

Interestingly, the origin of this clash is not related to the backreaction of inhomogeneities, but instead it is related to a violation of the strong equivalence principle. Namely,  some of these theories couple differently to gravitational energy with respect to other types of energy, but their naive cosmological versions do not capture that feature.  In order to illustrate this, we begin by imagining a simplified version of our universe.

\textbf{A vacuum universe:} We can conceive a ``vacuum'' universe formed only out of gravity. To this end, let us for the time being  consider Einstein's equations in the vacuum
\begin{equation}\label{EFE}
G_{\mu\nu}=0\, ,
\end{equation}
where for simplicity we are setting the cosmological constant to zero, but a nonzero $\Lambda$ would not change our discussion. We take this as our fundamental theory of gravity and we wish to study its implications for the large scale evolution of the universe.  
Obviously, a solution of \req{EFE} is simply the Minkowski spacetime but certainly not the only one. A black hole is also a solution, and more generally, a system of randomly distributed black holes, moving under their mutual gravitational influence, is also an exact solution.\footnote{There has been recent progress in numerically simulating this kind of universes in cosmology, which have been called ``black hole lattices" \cite{Bentivegna:2018koh}.} On top of this, we can also include a stochastic background of gravitational waves, which also solve the Einstein field equations in the vacuum. Let us note that the Ricci curvature of this solution is identically vanishing, while all the dynamics is encoded in the Weyl curvature. 

Now, over very large scales, this distribution of black holes and gravitational waves would look like a cosmic fluid with certain energy density and pressure.  Since the distribution is homogeneous and isotropic, at large scales one would describe this universe by a FLRW metric \req{FLRW}.  Naturally, this FLRW metric satisfies Einstein equations with a source
\begin{equation}\label{avEFE}
\hat{G}_{\mu\nu}=8\pi G\, \hat T_{\mu\nu}\, ,
\end{equation}
where $\hat{G}_{\mu\nu}$ is the Einstein tensor of the FLRW metric and $\hat T_{\mu\nu}$ is an effective stress-energy tensor corresponding to a perfect fluid. These equations applied to \req{FLRW} lead to the Friedmann equation
\begin{equation}\label{Friedmann}
H^2=\frac{8\pi G}{3}\rho\, ,
\end{equation}
where $H=\dot{a}/a$ and $\rho=\hat T_{tt}$ is the average energy density. 
Observe that, since we did not have any matter in our starting equation \req{EFE}, this energy density is in fact \emph{gravitational energy} in the form of black holes and gravitational waves. Indeed, the effective stress-energy tensor in \req{avEFE} emerges due to the fact that, in GR, gravity couples to itself. While in \req{EFE} we do not see this energy, which is encoded in the nonlinear form of the Einstein's equations, the averaging process makes it manifest in \req{avEFE}. This is the reason why, even in our own universe, the energy density of matter and radiation contains as well a gravitational contribution in the form of black holes and gravitational waves. 

We therefore have these two dual descriptions of the same universe:
\begin{itemize}
\item Fundamental description: the spacetime is Ricci flat, there is no matter and all the dynamics of gravity is encoded in the Weyl curvature. 
\item Cosmological description: the geometry is described by FLRW metrics sourced by an emergent stress-energy tensor. Ricci curvature is non-trivial due to this, while Weyl curvature is vanishing. 
\end{itemize}
It is particularly interesting that the averaging procedure somehow swaps Weyl and Ricci curvature, which makes it clear that the metrics in each description are different objects.

Now, we remark that it is non-trivial to establish the validity of \req{avEFE} starting from \req{EFE}, and this has been the subject of intensive research \cite{Zalaletdinov:1996aj,Boersma:1997yt,Coley:2005ei,Buchert:2007ik,Green:2010qy,Green:2014aga,Bolejko:2016qku}. In fact, \req{avEFE} would contain additional terms in the right-hand-side on account of the backreaction of inhomogeneities. However, those additional terms have been shown to be small \cite{Green:2014aga}, so that \req{avEFE} is indeed a good approximation to describe the average evolution of the universe.\footnote{There is still some debate about whether backreaction effects can become important \cite{Buchert:2015iva}. However, we do not wish to go into this question as the issue that we raise in this paper is not related to backreaction.} 
This is what justifies the application of Einstein's equations together with a perfect-fluid stress-energy tensor to describe cosmology. 
But, can we say the same about modifications of GR?

\textbf{Cosmological versus fundamental theories:}
As a consequence of the previous observations, it should already become clear that, if we write down our favorite theory of gravity, it is not the same to apply that theory to the effective cosmological metric as applying it to the fundamental metric and then deriving the consequences for cosmology, understood as an average. These two approaches will in general lead to inconsistent answers --- see \cite{Clifton:2012ry,Clifton:2015ira} for some explicit examples. Here we point out a new type of inconsistency that, as we argue, applies to a broad class of theories. 

Let us illustrate this in the case of the well-studied $f(R)$ gravity, with an action 
\begin{equation}\label{eq:f(R)}
S_{f(R)}=\frac{1}{16\pi G}\int d^{4}x\sqrt{-g}f(R)\, ,
\end{equation}
and the additional assumption that it is minimally coupled to matter. Thus, the theory has equations of motion 
\begin{equation}\label{f(R)eom}
\begin{aligned}
&f'(R)R_{\mu\nu}-\frac{1}{2}f(R) g_{\mu\nu}\\
&-\left(\nabla_{\mu}\nabla_{\nu}-g_{\mu\nu}\Box \right)f'(R)=8\pi G\,  T_{\mu\nu}\, .
\end{aligned}
\end{equation}
Let us then study the consequences of this theory for the vacuum universe we just constructed.

We first consider the ``fundamental'' point of view in which this universe is a soup of black holes and gravitational waves, and assume that the equations \req{f(R)eom} govern gravity at a fundamental level. Then, since we have no matter, we set $T_{\mu\nu}=0$ in  \req{f(R)eom}. Now, our original vacuum universe in Einstein gravity had $R_{\mu\nu}=0$, but this is also a solution of $f(R)$ gravity, since all Ricci flat metrics remain exact solutions of \req{f(R)eom} with $T_{\mu\nu}=0$.\footnote{For this we only need to assume that $f(R)$ is differentiable at $R=0$ and that we do not have a cosmological constant. If we had a nonzero cosmological constant, $f(R)$ gravity also preserves all the solutions of vacuum Einstein gravity, up to a renormalization of the cosmological constant.} One can even prove that black hole solutions must be given by the GR solutions for reasonable choices of $f(R)$ \cite{Whitt:1984pd,Mignemi:1991wa}. Therefore, $f(R)$ gravity would in principle leave unaffected this vacuum universe. At the very least, we can say that it does not necessarily introduces modifications to this solution. Obviously this also means that the large-scale evolution of the universe --- which arises upon averaging this solution over large distances --- remains unchanged and hence it is described by the usual Friedmann equation \req{Friedmann}. 
Thus, from the fundamental perspective, $f(R)$ gravity has no effect whatsoever on our vacuum universe. 

Let us then consider the other perspective and treat $f(R)$ gravity as a cosmological theory. Thus we directly apply the equations \req{f(R)eom} to FLRW metrics. This means that we put a hat on top of every quantity of \req{f(R)eom} (denoting that they are evaluated on cosmological metrics) and we also include a perfect-fluid stress-energy tensor $\hat T_{\mu\nu}$. This contains the average energy density and pressure of the cosmic fluid. In the case of our vacuum universe, its origin would be purely gravitational,  just like in \req{avEFE}.
We remark that this is the standard way in which $f(R)$ gravity and other theories are used for cosmological applications in the literature: one simply includes a perfect-fluid stress-energy tensor in the right-hand side of the equations of motion.  

In this approach, one can see that  $f(R)$ gravity indeed leads to consequences for cosmology with respect to Einstein gravity. 
For instance, if $f(R)=R+\frac{\alpha}{12}R^2$, then the first Friedmann equation becomes
\begin{equation}\label{FriedmannR2}
H^2=2\alpha\left[\Psi(\Psi-H^2)-H\dot\Psi\right]+\frac{8\pi G}{3}\rho\, ,
\end{equation}
where $\Psi=H^2+\frac{1}{2}\dot H$. In general, $\Psi$ provides a new degree of freedom that cannot be set to zero, and it affects the evolution of the scale factor. One can see that, for a universe that contains a matter density --- which behaves as $\rho_{\rm m}\propto a^{-3}$ --- the evolution is \emph{necessarily} modified with respect to GR on account of \req{FriedmannR2} (the term proportional to $\alpha$ cannot vanish). This would include the case of a universe containing only black holes, which act like a matter density when averaged over large distances. However, we just saw that from the fundamental perspective, $f(R)$ gravity does not modify this kind of universe. Therefore we have reached a contradiction. Certainly, this must come from the naive expectation that one can apply \req{f(R)eom} directly in the cosmological setup. Somehow, the averaging process must change these equations in a fundamental way.

\textbf{Adding matter:} In order to get a deeper understanding of the reason why cosmological $f(R)$ gravity is inconsistent, we can now consider a universe where we have usual matter and radiation, along with gravitational ``matter'' and radiation in the form of black holes and gravitational waves. 
While $f(R)$ theories do not modify black hole solutions  --- or any other Ricci-flat metric --- they do affect the gravitational field of matter distributions (in particular, the exterior field of a spherically symmetric body is not described by the Schwarzschild metric \cite{Bueno:2017sui,Casado-Turrion:2023rni}). This is because non-gravitational energy acts as a source for the scalar mode propagated by $f(R)$ gravity, while this mode is inert to gravitational energy. 

Thus, in the presence of actual matter one can expect that $f(R)$ gravity indeed leads to a modification of the evolution of the universe. However, the application of \req{f(R)eom} to cosmological metrics would again be inconsistent, because $\hat{T}_{\mu\nu}$ would contain all types of energy, also the gravitational one. While the fundamental $f(R)$ theory distinguishes between gravitational and non-gravitational energies, its cosmological version  couples universally to all forms or energy. 
In particular, the equation \req{FriedmannR2}, which does not discriminate between gravitational or non-gravitational energy density, cannot arise from the macroscopic average of $f(R)$ gravity and must be fundamentally wrong. 

This is better understood if we transform $f(R)$ gravity into the Einstein frame, in which case it becomes Einstein gravity plus a minimally coupled scalar field
\begin{equation}
S=\frac{1}{16\pi G}\int d^{4}x\sqrt{|g|}\left[R-\frac{1}{2}(\partial\phi)^2-V(\phi)\right]+S_{\rm m}\, .
\end{equation}
The scalar field couples to the trace of the stress energy tensor giving rise to equations of motion of the form
\begin{align}
G_{\mu\nu}-\frac{1}{2}\partial_{\mu}\phi\partial_{\nu}\phi-\frac{1}{2}g_{\mu\nu}\mathcal{L}(\phi)&=8\pi G\, T_{\mu\nu}^{\rm matter}\, ,\\
\nabla^2\phi-V'(\phi)&=h(\phi) T^{\rm matter}\, ,
\end{align}
where $\mathcal{L}(\phi)=-\frac{1}{2}(\partial\phi)^2-V(\phi)$ and $h(\phi)$ is a function that determines the coupling of the scalar to the matter. 

However, the cosmological averaged equations would give something like
\begin{align}
\hat{G}_{\mu\nu}-\frac{1}{2}\partial_{\mu}\hat\phi\partial_{\nu}\hat\phi-\frac{1}{2}g_{\mu\nu}\mathcal{L}(\hat\phi)&=8\pi G\, \hat{T}_{\mu\nu}^{\rm eff}\, ,\\
\nabla^2\hat{\phi}-V'(\hat{\phi})&=h(\hat\phi) \hat{T}^{\rm matter}\, ,
\end{align}
where now the effective stress-energy tensor in Einstein field equations contains as well a gravitational contribution, 
\begin{equation}
\hat{T}_{\mu\nu}^{\rm eff}=\hat{T}_{\mu\nu}^{\rm grav}+\hat{T}_{\mu\nu}^{\rm matter}\, ,
\end{equation}
which appears for the same reason we got a nonzero $\hat{T}_{\mu\nu}$ in \req{avEFE}. Note however, that the in the scalar equation we still have the averaged matter stress-energy tensor, but not the total effective stress energy tensor. 

Thus, if we have a universe where all the energy content is gravitational --- in the form of black holes and gravitational waves --- then $\hat{T}_{\mu\nu}^{\rm matter}=0$, the scalar field is not active and the evolution corresponds to the standard Friedmann equation. However, if we have actual matter, then the scalar field is excited and it leads to modifications from GR. In general, we have an intermediate situation where the scalar field is active but the energy density that sources it is not the same energy density appearing in the Friedmann equations. 
We remark that this is not the usual way in which this theory is treated in the literature, where a distinction between the types of energy is not considered. Furthermore, this distinction is problematic in the original formulation \req{eq:f(R)}, because in that case the scalar degree of freedom is nonminimally coupled. 

\textbf{Other theories:} 
Similar comments apply to many other modified gravity theories.  For instance, we can generalize the previous discussion to any theory built from the Ricci tensor, $\mathcal{L}=f(R_{\mu\nu})$. All these theories admit Einstein spacetimes as exact solutions, so they do not induce modifications to a vacuum universe consisting of black holes and gravitational waves. However, they yield nontrivial effects if treated as cosmological theories, \textit{i.e.}, if we evaluate their equations of motion on an FLRW metric and add a perfect-fluid stress energy tensor. Thus, we have the same kind of inconsistency between both approaches. 
We expect the same problem in any theory with extra gravitational degrees of freedom that couple universally to matter, like in the case of Horndeski theories. In all those cases, the extra degrees of freedom are sourced by explicit matter energy but not by gravitational energy, and therefore the standard application of those theories in cosmology --- where one assumes that the stress-energy tensor that enters into Einstein field equations is the same one that sources the extra degrees of freedom --- is inconsistent. In some cases one could fix this problem by reformulating the theory in a way that it can distinguish between the different types of energy, but this may not always be possible in all cases, especially if the extra degrees of freedom are nonminimally coupled.

\textbf{Discussion:}
We have found a clash in the naive application of some theories of modified gravity in the cosmological context. The reason of the clash is the following: some of these theories contain extra degrees of freedom that couple differently to gravitational and non-gravitational energy. This is very clear in the case of $f(R)$ gravity, where we have a scalar mode that is not sourced by gravitational energy (\textit{e.g.}, black holes), but it is sourced by usual matter with an explicit stress-energy tensor. However, in the standard application of these theories in cosmology one includes an effective stress-energy tensor that, by construction, contains all types of energy, also gravitational one. This leads to inconsistent answers depending on how one interprets the theory: as an effective cosmological theory, or as a fundamental theory from which one extracts the cosmological dynamics by averaging over large distances. The contradiction is most evident when we consider a universe where all the ``matter'' content is gravitational (black holes and gravitational waves): from a fundamental perspective these theories do not modify any prediction of GR, but in a direct cosmological approach they do. We remark that this effect would be relevant for the application of modified gravity theories to our own universe if, for instance, a large fraction of dark matter turned out to be black holes. 

Now, let us be clear about the implications of this result. This inconsistency does not mean that one cannot use these theories as cosmological models. However, it means that the origin of these models as the average of a fundamental theory is unclear. In fact, a relevant question would be whether for any cosmological modified gravity there is a fundamental theory of gravity for which the former arises as a macroscopic average description. We have shown in this letter that cosmological $f(R)$ gravity cannot come from the average of the same (or actually any) $f(R)$ theory, and similarly for other related theories.  We have no hint at this point of what that theory could be --- if it exists at all.  Our result implies anyway that one cannot use one of these theories simultaneously to explain cosmology and to explain gravity as a fundamental interaction.

Establishing the map from fundamental to cosmological theories is indeed a challenging question that we need to address if we wish to perform tests of fundamental physics with cosmology. Bringing attention to the relevance of this poorly understood problem was one of the main goals of this letter. 

In fact, our findings may also have implications for theories that are traditionally ignored for cosmological applications. This is the case of the effective field theory (EFT) of GR \cite{Endlich:2017tqa,Cano:2019ore,Ruhdorfer:2019qmk}, which captures general corrections to vacuum Einstein gravity. In this theory one includes higher-derivative terms formed from Weyl curvature, and as such they are irrelevant for FLRW metrics which are conformally flat. However, when one understands cosmology as an average theory, those terms could induce modifications. EFT corrections indeed introduce modifications to black hole solutions, and by extension, they also affect a universe filled with black holes, potentially affecting the large scale evolution.  In this way, EFT modifications of GR, which are naturally expected from high-energy physics, could unexpectedly be relevant for cosmology --- especially during the very early universe. Establishing the cosmological theory arising from these EFT corrections is an interesting problem.

\vspace{0.1cm}
\begin{acknowledgments} 
I would like to thank Pablo Bueno, Marina David, Jaume Garriga, Robie Hennigar, Tom\'as Ort\'in, Alejandro Ruip\'erez and especially Jos\'e Beltr\'an for useful comments and discussions. 
The work of PAC received the support of a fellowship from “la Caixa” Foundation (ID 100010434) with code LCF/BQ/PI23/11970032. 
\end{acknowledgments}

\bibliography{StringGravity}
\noindent \centering

\end{document}
%